# Comparison Between Damping Coefficients of Measured Perforated Structures and Compact Models

T. Veijola[1], G. De Pasquale[2], and A. Somá[2]

[1] Department of Radio Science and Engineering, Helsinky University of Technology
P.O. Box 3000, 02015 TKK, Finland.

[2] Department of Mechanics, Polytechnic of Torino
Corso Duca degli Abruzzi 24, 10129 Torino, Italy.

*Abstract* - Measured damping coefficients of six different perforated micromechanical test structures are compared with damping coefficients given by published compact models. The motion of the perforated plates is almost translational, the surface shape is rectangular, and the perforation is uniform validating the assumptions made for compact models. In the structures, the perforation ratio varies from 24% - 59%. The study of the structure shows that the compressibility and inertia do not contribute to the damping at the frequencies used (130kHz - 220kHz). The damping coefficients given by all four compact models underestimate the measured damping coefficient by approximately 20%. The reasons for this underestimation are discussed by studying the various flow components in the models.

## I. INTRODUCTION

Perforations are used in micromechanical squeeze-film dampers for several reasons. The main purpose is to reduce the damping and spring forces of oscillating structures due to the gas flow in small air gaps. Generally, the modeling problem is quite complicated, since the damping force acting on the moving structure depends on the 3D fluid flow in the perforations, in the air gap, and also around the structure. Compact models have been published in the literature, but their verification is generally questionable. Verification methods used are FEM solutions of the Navier-Stokes equations of the fluid volume and measurements [1].

In this paper, responses of four compact models are compared with measured responses of test structures. Six different perforated plates (figures 1 and 2) with different topologies have been measured at their first out-of-plane resonant frequencies, and the damping coefficients have been calculated from the quality factors (Q values) and effective masses. The measurement setup, the testing procedure and specimens characteristics are presented in [2]; here it is observed that dynamic parameters of the microsystem characterizing the fluidic and structural coupling can be extracted from the experimental frequency response function (FRF). The dynamic performance of microstructures are discussed based on the analytical solutions to perforated parallel-plate problems in [3], [4] and [5]. Since the perforation is uniform, the motion is almost translational, and also since the shape of the surface is rectangular, analytic damping models are applicable.

First, the oscillating flow is analyzed using several characteristic numbers, the applicable modeling method is then chosen, the damping coefficients are calculated and compared with the measured ones. Finally, the results are discussed.

## II. TEST STRUCTURES AND MEASUREMENTS

Figures 1 and 2 show the structures of the test specimen. The height of the plate $h_c$ = 15μm, the air gap height $h$ = 1.6μm. Table I shows the other dimensions of the measured devices. In the table, $q$ is the perforation ratio in percent, in this case $q = MNs_0^2/(LW)$, $M$ and $N$ are the number of holes in the length and width directions, respectively.

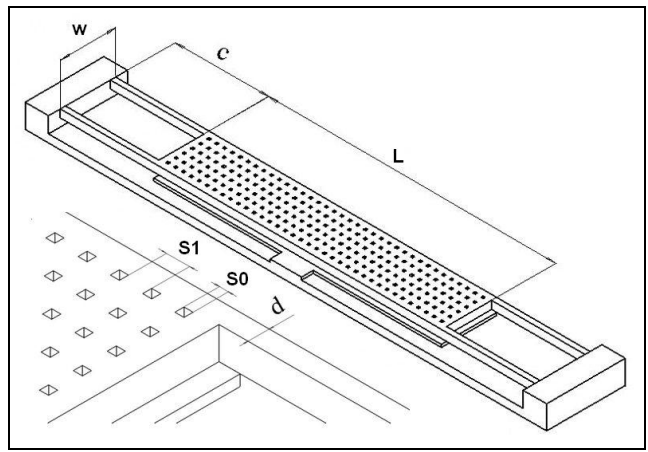

Fig. 1. Geometrical shape and dimensions of the vibrating structures.

The measurements are made using the interferometric microscope Fogale ZoomSurf 3D (figure 3), with 20x objective magnification factor, 0.1nm of vertical resolution and 0.6μm of lateral resolution.

The *Frequency Shift* technique is used, consisting of the excitation of the structure by an alternate voltage, the frequency of which is progressively increased by discrete steps. For each level of actuation frequency, the corresponding amplitude of vibration is stored and the experimental FRF is plotted for the detection of the resonance peak (figure 4). The first detection is made across a wide frequency range (0-500kHz) in order to roughly locate the resonance; five successive identical detections are then performed across a more precise and narrow range. These are statistically treated to extract the values of resonance reported in Table II.

TABLE I
DIMENSIONS OF MEASURED TOPOLOGIES

| type | $L$ [μm] | $W$ [μm] | $M \times N$ | $L:W$ | $s_0$ [μm] | $s_1$ [μm] | $q$ % |
|---|---|---|---|---|---|---|---|
| A | 372.4 | 66.4 | 36x6 | 6:1 | 5.0 | 5.2 | 24 |
| B | 363.9 | 63.9 | 36x6 | 6:1 | 6.1 | 3.9 | 37 |
| C | 373.8 | 64.8 | 36x6 | 6:1 | 7.3 | 3.0 | 50 |
| D | 369.5 | 64.5 | 36x6 | 6:1 | 7.9 | 2.3 | 59 |
| E | 363.8 | 123.8 | 36x12 | 3:1 | 6.2 | 3.8 | 38 |
| F | 363.8 | 243.8 | 36x24 | 3:2 | 6.2 | 3.8 | 38 |

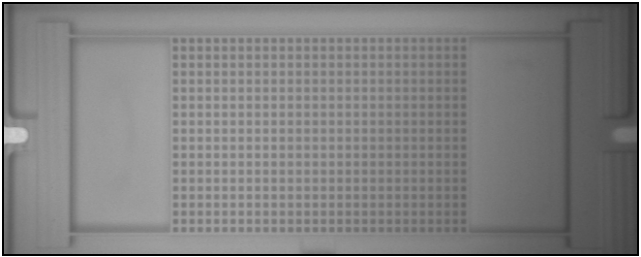

Fig. 2. Microscope image of specimen F.

The measurement technique described uses a red monochromatic light source for the interferometric fringes detection. The vibration amplitude is detected optically in correspondence of a selectable region (*detection window*) of the specimen, located at the center of the suspended plate. The output value of the oscillation amplitude is averaged between the values captured by each pixel of the CCD camera inside the active window.

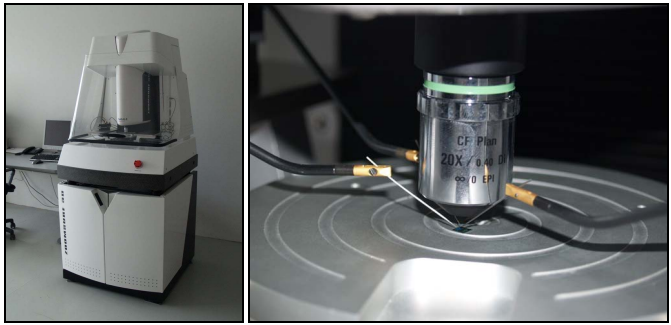

Fig. 3. Interferometric microscope Fogale ZoomSurf 3D (a) and the 20x Nikon objective (b).

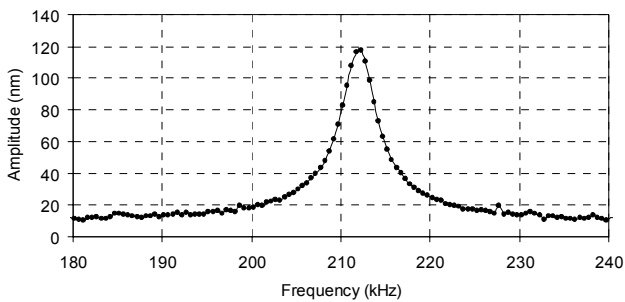

Fig. 4. Displacement vs. frequency diagram of specimen C.

The quality factor is extracted from the experimental curve, that was previously interpolated by a 6-order polynomial; the damping coefficient is finally calculated from the quality factor, resonant frequency and the effective mass by means of the method of the *half power bandwidth*. These are shown in Table II. The effective mass is calculated from FEM eigenmode analysis. The mass ratio α is the ratio between the modal mass and total mass.

TABLE II
MEASURED DAMPING COEFFICIENTS AND RESONANT FREQUENCIES OF SIX DIFFERENT TEST STRUCTURES

| type | $c_m$ measured [$10^{-6}$Ns/m] | $f_0$ measured [kHz] | mass ratio α |
|---|---|---|---|
| A | 47.38 | 201.637 | 0.918 |
| B | 19.46 | 204.329 | 0.893 |
| C | 9.863 | 211.011 | 0.885 |
| D | 7.609 | 222.282 | 0.856 |
| E | 38.22 | 173.904 | 0.946 |
| F | 67.44 | 138.564 | 0.974 |

It is ensured that the amplitude of the oscillation is small compared with the gap height, and the static bias voltage caused by the excitation signal and DC bias voltage does not deflect the plate changing the air gap height.

III. MODELING OF THE DAMPING COEFFICIENT

A. *Analysis of the fluid flow*

The actual mass is supported with thin beams in such a way that the movement of the mass is approximately translational. This justifies starting with an analysis where the velocity of the plate surface is constant.

*Flow patterns*

In perforated dampers in perpendicular motion, two different flow patterns can be distinguished. The first is the "closed borders" pattern, where the fluid flows only through the holes. The second pattern, the "closed holes" pattern considers only the flow from the damper borders. In practical dampers both patterns exist simultaneously. The perforation ratios $q$ (the area of the surface without holes divided by the area of the holes) are here considerably high, ranging from 24% - 59%. Also, the holes are relatively wide compared to the air gap height. It is then evident that the first "closed borders" flow pattern is strong here. Here compact models that consider both flow patterns are selected; the contribution the flow patterns in the measured cases will be discussed later in this paper.

*Rarefied gas*

Air at standard atmospheric conditions is used in the measurements. The pressure $P_A$ = 101kPa, the density ρ = 1.155 kg/m$^3$, the viscosity coefficient μ = 18.5·10$^{-6}$ Ns/m$^2$, and the mean free path λ = 65nm. The air gap height $h$ = 1.6μm, which makes the Knudsen number of the air gap flow $K_{ch} = \lambda/h$ to be 0.04, and the smallest hole diameter is 5μm making the Knudsen number of the perforation $K_{tb} = \lambda/s_0$ to be 0.013. The estimated contributions to the damping coefficient are $1/(1 + 6K_{ch})$ and $1/(1 + 7.567K_{tb})$, that is -24% and -9.8%, respectively. The assumption of a slightly rarefied gas is justified, and the slip velocity model is sufficient.

*Compressibility*

Next, the contribution of compressibility is analyzed. This can be made by studying the squeeze number. For a rigid surface, the squeeze number is

$$\sigma = \frac{12\eta W^2 \omega}{P_A h^2} \quad (1)$$

$W$ is here the smallest, "dominating" characteristic dimension. For example, for case A without holes

$$\sigma = 3.8 \cdot 10^{-6} \omega \quad (2)$$

At 200kHz $\sigma$ = 4.8. This means that without perforation, compressibility should be considered (when $\sigma \sim 20$, the viscous and spring forces are equal). When the surface is perforated, the situation changes completely. The characteristic dimension can now be estimated to be the space between the holes. In case A this is 5.2μm. The squeeze number now becomes $\sigma$ = 0.03 at 200kHz. This is an overestimate of the real situation, since accounting for the damping in the holes will make the effective squeeze number considerably smaller.

According to the squeeze number analysis, the spring forces are negligible compared with the damping forces, and noncompressible gas can be assumed without loss of any accuracy. The damping coefficient can be considered constant up to several MHz.

Also, it is expected that the spring force due to the gas is much smaller than the force due to the effective spring of the mechanical structure. The frequency shift due to gas compressibility is expected to be very small. The spring force in the system is only due the effective spring of the mass-spring system. The spring coefficient can also be considered constant at least up to several MHz.

*Gas Inertia*

One can suspect that the inertia of the gas may contribute to the damping coefficient. The place where the inertia is the largest is the "widest" flow channel, that is the perforation holes.

The contribution of inertia is characterized by the Reynolds number $R_e$, specified for a circular channel as [6]:

$$R_e = \frac{\rho r^2 \omega}{\mu} \quad (3)$$

The "worst" case, where the inertial is the largest, is case D where the hole diameter is the largest. If the square channel is approximated with a circular channel having a radius of $r = 4\mu m$, that's half of the hole side, gives $R_e = 0.998 \cdot 10^{-6} \omega$, and at 200 kHz $R_e$ = 1.255. The real and imaginary parts of the impedance are equal when $R_e \sim 6$. The additional imaginary part does not directly influence the damping coefficient, but the change is due to the frequency dependent real part. At $R_e \sim 6$, the change in the real part in only 3.2% [6].

This Reynolds number study shows that the inertia needs not to be considered even in the accurate analysis.

*B. Compact models*

A model for the noncompressible perforation cell is sufficient, as indicated in the study above. Four models that consider both "closed holes" and "closed borders" flow patterns are selected to be compared. The first one, M1, is a model by Bao [3] for a rectangular damper that is much longer than wide. In this model the air gap regime flow resistance, the flow resistance of a circular perforation, and its constant elongation are included. Continuum flow conditions are assumed. The 2nd model M2 has been also presented in [3], but now an arbitrary rectangular surface is assumed. Next, the model M3 in [4] is used. The air gap flow resistance model, the circular perforation flow channel model, and four different elongations of the flow channels, that vary depending on the ratios of the cell dimensions, are included in the model. Slip velocity conditions are used for the air gap and the perforations. The 4th model, M4, is made especially for square holes [5]. It includes similar components as the previous model and accounts also for the rarefied gas in the slip flow regime.

Model M4 for square perforations accepts directly the dimensions given in Table I (size of the perforation cell $s_x = s_0 + s_1$). To apply the other models, an effective radius of the circular perforation $r_0$ and the cell $r_x$ need to be specified first. Matching the areas of the actual cell $s_x^2$, and the equivalent circular cell $\pi r_0^2$ gives

$$r_x = \frac{s_x}{\sqrt{\pi}} \quad (4)$$

The radius $r_0$ is determined by requiring the acoustic impedances of square and rectangular channels to match. For relatively small Knudsen numbers this leads to approximately [3], [4]

$$r_0 = \frac{1.096 s_0}{2} \approx 0.547 s_0 \quad (5)$$

The Appendix shows all equations needed in computing the damping coefficients using models M1…M4.

IV. SIMULATION RESULTS AND DISCUSSION

The results of the comparison are shown in Table III. $\Delta_i$ is the relative error of the simulated damping coefficient $c_s$ of model M$i$ compared to the measured damping coefficients.

TABLE III
RELATIVE ERRORS OF THE COMPACT MODELS

| type | $\Delta_1$ [%] M1 | $\Delta_2$ [%] M2 | $\Delta_3$ [%] M3 | $\Delta_4$ [%] M4 |
|---|---|---|---|---|
| A | -23.53 | -25.74 | -33.51 | -33.27 |
| B | -16.36 | -18.06 | -21.02 | -21.96 |
| C | -5.21 | -6.59 | -4.11 | -6.65 |
| D | -14.66 | -15.72 | -12.46 | -15.29 |
| E | -17.27 | -18.94 | -19.03 | -20.14 |
| F | -4.77 | -6.70 | -5.19 | -6.52 |

The results of models M3 and M4 are quite close to each other, the largest error between them is only 2.8%-points, showing that the effective radius approach is sufficient. Continuum conditions were assumed for M1 and M2, giving approximately 10% larger values than with slip velocity conditions. If these models are corrected to account for the rarefied gas, the errors become approximately 10% worse, than those shown in Table III.

The drag on the sidewalls is expected to increase the damping coefficient since the structure is relatively high. The models do not account for this drag force. The moving supporting beams will have also an additional contribution to the damping. The length of the supporting beams in all cases is about $L_b$ = 122μm, and their widths are about $W_b$ = 4μm. A rough approximation for the contribution of the supporting beams is

$$c_b = \frac{4}{3}\frac{L_b(W_b+1.3h)^3 \mu}{h^3(1+6K_{ch})} \qquad (6)$$

This gives a damping coefficient of $c_b$ = 0.16·10$^{-6}$ Ns/m. This approximation shows that the damping due to the beams is very small.

The responses of models M1 and M2 are quite close. One could expect that the error of M1 would be larger in case E and especially in case F, since the length to width ratio is quite small in these cases. The explanation for this can be found by studying the contribution of the different flow patterns. This can be easily done using the perforation cell model that assumes the "closed borders" flow pattern, where the pressure distribution is independent of the shape of the damper. The flow resistances $R_P$ for a perforation cell in [4] and [5] are derived using this assumption. The damping coefficient becomes simply $c_P = NMR_P$ ($MN$ is the number of holes). Table IV shows the errors of the damping coefficients $c_P$ compared with the measured values using the models for circular cells, M5, and rectangular cells, M6, as presented in [4] and [5], respectively.

TABLE IV
RELATIVE ERROR OF "PERFORATION CELL" COMPACT MODELS

| type | $\Delta_5$ [%] M5 | $\Delta_6$ [%] M6 |
|---|---|---|
| A | -17.25 | -16.92 |
| B | -7.81 | -9.00 |
| C | 7.38 | 4.36 |
| D | -3.55 | -6.83 |
| E | -11.45 | -12.73 |
| F | 0.37 | -1.08 |

The results in Table IV show that the "closed borders" flow pattern is the dominant one. The contribution of the "closed holes" flow is only 6% - 16% of the damping coefficient. This explains why models M1 and M2 differ only slightly in this case: the contribution of the shape-dependent damping is quite small.

TABLE V
RELATIVE CONTRIBUTIONS OF THE FLOW RESISTANCES OF MODEL M5

| type | $R_s$ [%] | $R_{is}$ [%] | $R_{ib}$ [%] | $R_{ic}$ [%] | $R_c$ [%] | $R_e$ [%] |
|---|---|---|---|---|---|---|
| A | 8.15 | 9.78 | 0.78 | 5.63 | 68.01 | 7.65 |
| B | 7.62 | 12.94 | 1.87 | 5.13 | 64.05 | 8.40 |
| C | 6.48 | 15.30 | 3.50 | 4.55 | 61.19 | 8.98 |
| D | 4.51 | 14.49 | 5.03 | 4.06 | 62.59 | 9.31 |
| E | 7.51 | 13.18 | 2.00 | 5.07 | 63.80 | 8.45 |
| F | 7.51 | 13.18 | 2.00 | 5.07 | 63.80 | 8.45 |

To study further the sources of damping, Table V shows the contributions of the flow resistance components in M5 [4]. The flow resistance of the perforations $R_S$ is the most significant source of damping; its contribution is approximately 65%. The 2$^{nd}$ important contribution comes from the intermediate region resistances $R_{IS}$, $R_{IB}$, and $R_{IC}$: 15 - 20%. Next important is the elongation at the perforation outlet $R_E$, about 8%.

V. CONCLUSIONS

Measured damping coefficients have been compared to those obtained with four different compact models for perforated dampers. After analyzing the oscillating flow with several characteristic numbers, sufficient models were selected. Only translational motion was assumed. The results of all models were quite close to each other, a systematic underestimate of the damping coefficient was about -20%. The reasons for this were discussed and the contribution of various flow components were presented. For a more accurate analysis, the realistic modes of the plates should be considered. It is expected that the "closed holes" flow pattern will become relatively stronger in this case.

The comparison showed also how a model for circular perforations can be used to model square holes.

APPENDIX

This Appendix contains equations for four compact models M1…M4. The dimensions and symbols in Fig. 1 are used: the length and width of the perforated plate are $L$ and $W$. The side lengths of the square holes and the square perforation cells are $s_0$ and $s_X = s_0 + s_1$, respectively.

*A.  Model M1 equations*

The equations for a narrow hole plate ($L>>W$) are given in [3]. Note, in the following equations $a = W/2$ and $b = L/2$. The equivalent radii for the circular cell and hole are given in Eqs. (4) and (5). The damping coefficient c is

$$c = 2aL \frac{8\mu h_C}{\beta^2 r_0^2}\left(1 + \frac{3r_0^4 K(\beta)}{16 h_C h^3}\right)\left[1 - \frac{l}{a}\tanh\left(\frac{a}{l}\right)\right]$$

where

$$K(\beta) = 4\beta^2 - \beta^4 - 4\ln\beta - 3$$

$$l = \sqrt{\frac{2h^3 H_{eff}\eta(\beta)}{3\beta^2 r_0^2}}$$

$$\eta(\beta) = 1 + \frac{3r_0^4 K(\beta)}{16 h_C h^3}$$

$$\beta = \frac{r_0}{r_C}$$

$$H_{eff} = h_C + \frac{3\pi r_0}{8}$$

*B.  Model M2 equations*

The equations for an arbitrary shaped rectangular plate are also included in [3]. Also, in the following equations $a = W/2$ and $b = L/2$. The equivalent radii for the circular cell and hole are given in Eqs. (4) and (5). The damping coefficient c is

$$c = -\gamma \frac{\mu(2a)^3(2b)}{h^3}$$

where

$$\gamma = 3\alpha^2 - 6\alpha^3 \frac{\sinh^2(1/\alpha)}{\sinh(2/\alpha)} - \frac{24\alpha^3\kappa}{\pi^2}\sum_{n=1,3,5,\ldots}^\infty \frac{\tanh\frac{\sqrt{1+(n\pi\alpha/2)^2}}{\alpha\kappa}}{n^2\left[1+(n\pi\alpha/2)^2\right]^{\frac{3}{2}}}$$

$$\kappa = \frac{a}{b}, \quad \alpha = \frac{l}{a}$$

Above, $l$ is the same as used in M1 equations.

*C.  Model M3 equations*

A model for a circular perforation cell is derived in [4], and the damping coefficient of a rectangular perforated plate is given in the paper. Note, in the following equations $a = W$ and $b = L$. The equivalent radii for the circular cell and hole are given in Eqs. (4) and (5). The damping coefficient c is

$$c = \sum_{m=1,3,5,\ldots}^\infty \sum_{n=1,3,5,\ldots}^\infty \frac{1}{G_{m,n}(a_{eff}, b_{eff}) + 1/R_{m,n}} \quad (C1)$$

Where the effective surface dimensions are

$$a_{eff} = a + 1.3(1 + 3.3K_{ch})h$$
$$b_{eff} = b + 1.3(1 + 3.3K_{ch})h$$

and

$$G_{m,n}(a,b) = \left(\frac{m^2}{a^2} + \frac{n^2}{b^2}\right)\frac{m^2 n^2 \pi^6 h^3 Q_{ch}}{768\mu ab}$$

$$R_{m,n} = \frac{64 MN R_P}{m^2 n^2 \pi^4}$$

The flow resistance of a single perforation cell is

$$R_P = R_S + R_{IS} + R_{IB} + \frac{r_X^4}{r_0^4}(R_{IC} + R_C + R_E)$$

$$R_S = \frac{12\pi\mu r_X^4}{Q_{ch} h^3}\left(\frac{1}{2}\ln\frac{r_X}{r_0} - \frac{3}{8} + \frac{r_0^2}{2r_X^2} - \frac{r_0^4}{8r_X^4}\right)$$

$$R_{IS} = \frac{6\pi\mu(r_X^2 - r_0^2)^2}{r_0 h^2}\Delta_S$$

$$R_{IB} = 8\pi\mu r_0 \Delta_B$$

$$R_{IC} = 8\pi\mu r_0 \Delta_C$$

$$R_C + R_E = 8\pi\mu\left(\frac{h_C}{Q_{tb}} + \Delta_E r_0\right)$$

where the elongations are

$$\Delta_S = \frac{0.56 - 0.32\frac{r_0}{r_X} + 0.86\frac{r_0^2}{r_X^2}}{1 + 2.5 K_{ch}}$$

$$\Delta_B = 1.33\left(1 - 0.812\frac{r_0^2}{r_X^2}\right)\frac{1 + 0.732 K_{tb}}{1 + K_{ch}}f_B\left(\frac{r_0}{h}, \frac{h_C}{h}\right)$$

$$\Delta_C = (1 + 6K_{tb})\left(0.66 - 0.41\frac{r_0}{r_X} - 0.25\frac{r_0^2}{r_X^2}\right)$$

$$\Delta_E = \frac{0.944 \cdot 3\pi(1 + 0.216 K_{tb})}{16} \times \left(1 + 0.2\frac{r_0^2}{r_X^2} - 0.754\frac{r_0^4}{r_X^4}\right)f_E\left(\frac{r_0}{h}\right)$$

where the functions are

$$f_B(x, y) = 1 + \frac{x^4 y^3}{7.11(43y^3 + 1)}$$

$$f_E(x) = 1 + \frac{x^{3.5}}{178(1 + 17.5 K_{ch})}$$

The flow rate coefficients and Knudsen numbers for the air gap and the holes are

$$Q_{ch} = 1 + 6K_{ch}, \quad K_{ch} = \frac{\lambda}{h}$$

$$Q_{tb} = 1 + 4K_{tb}, \quad K_{tb} = \frac{\lambda}{r_0}$$

*D.   Model M4 equations*

A model for a rectangular perforation cell has been given in [5]. Note, in the following equations $a = W$ and $b = L$. The damping coefficient c is given by (C1), where $R_P$ for a rectangular hole is

$$R_P = R_S + R_{IS} + R_{IB} + \frac{s_X^4}{s_0^4}(R_{IC} + R_C + R_E)$$

$$R_S = \frac{12\pi\mu r_X^4}{Q_{ch}h^3}\left(\frac{1}{2}\ln\frac{r_X}{r_{0E}} - \frac{3}{8} + \frac{r_{0E}^2}{2r_X^2} - \frac{r_{0E}^4}{8r_X^4}\right)$$

$$R_{IS} = \frac{3\mu(s_X^2 - s_0^2)^2}{s_0 h^2}\Delta_S$$

$$R_{IB} = 0$$

$$R_{IC} = 28.454\mu s_0 \Delta_C$$

$$R_C + R_E = 28.454\mu\left(\frac{h_C}{Q_{sq}} + \Delta_E s_0\right)$$

where the elongations are

$$\Delta_S = 0.122(1 + 6.5\xi - 3.8\xi^2)$$

$$\Delta_C = 0.302$$

$$\Delta_E = 0.242(1 + 4K_{sq})(1 - \xi^4)\left(1 + 0.019\left(\frac{s_0}{h}\right)^{2.83}\right)$$

where

$$\xi = \frac{s_0}{s_X}$$

The equation for $\Delta_E$ includes a misprint in [5]. The corrected equation is shown above.

The flow rate coefficients and Knudsen numbers for the square hole are

$$Q_{sq} = 1 + 7.567K_{sq}, \quad K_{sq} = \frac{\lambda}{s_0}$$

The effective radius is

$$r_{0E} = \frac{0.58076 s_0}{1 + 0.02108\xi^2 + 0.008\xi^4}$$